# Numerical study on mechanism of C-J deflagration

**Yunfeng Liu**[1,2]*

1. Institute of Mechanics, Chinese Academy of Sciences, Beijing 100190, China

2. School of Engineering Science, University of Chinese Academy of Sciences, Beijing 100049, China

liuyunfeng@imech.ac.cn

**Abstract**: The mechanism of detonation instability and deflagration-to-detonation transition is studied by one-dimensional numerical simulation with overall one-step chemical reaction kinetics in this paper. The mixture is stochiometric $H_2$-air mixture at 1atm and 300K. The detonation is ignited at the left closed end of the one-dimensional detonation tube and propagates downstream. The activation energy is increased to trigger the instability of detonation. The numerical results show that the C-J detonation is stable at lower activation energy. The stable detonation does not have the von Neumann spike and the gas Mach number at detonation front is subsonic. The von Neumann spike appears and the gas Mach number becomes supersonic as the activation energy is increased. The detonation instability appears with the von Neumann spike synchronously. The mechanism of detonation instability can be explained by the linear stability theory of Rayleigh, which says that the velocity profile with an inflectional point is unstable for inviscid flow. At very higher activation energy, the detonation quenches abruptly and degenerates into a C-J deflagration. The detonation is extinguished abruptly by the rarefaction wave induced by the higher von Neumann spike, which is stronger than the heat release. Then the rarefaction wave moves in front of the heat release region and weakens the leading shock wave gradually. The C-J deflagration is composed of a precursor shock wave and a flame front, and the flame front is completely decoupled from the shock wave. The gas static temperature behind the leading shock wave is too low to ignite the mixture. The rarefaction wave from the wall ceases the mixture behind the leading shock, increases its static temperature and decrease its pressure. As a result, the combustion takes place at the interface. The pressure rise caused by the combustion at the interface offsets the influence of rarefaction wave, and this mechanism makes the C-J deflagration propagate downstream with a relatively constant velocity for a long time.

## 1. Introduction

In the deflagration-to-detonation transition (DDT) experiments, a quasi-steady detonation propagating with about half C-J detonation speed ($D_{C-J}$) was frequently observed, which is called quasi-detonation or C-J deflagration [1-16]. It is called C-J deflagration in this paper. The C-J deflagration can propagate with almost constant velocity for a certain long period before an abrupt transition to a detonation or quenches. This means that C-J deflagration is a critical condition for DDT process.

Chu et al. conducted two types of experiments about C-J deflagration [3-10]. The first is the C-J deflagration propagating in tubes containing obstacles, and the second is deflagration created from established detonation by eliminating the transverse waves by the absorbing walls. They found that the speed of C-J deflagration is close to the sound speed of detonation product. A theoretical analysis was carried out on a configuration that consists of a plane precursor shock-wave driven by a plane C-J deflagration. Their analytical result shows that the combination of shock and deflagration is controlled by the energetics of the reacting mixture.

Frolov reviewed the research about deflagration-to-detonation transition in gases in the Semenov Institute of Chemical Physics of the Russian Academy of Sciences [11,12]. Guo et al. carried out experimental investigation of the propagation of gaseous detonation waves over tube sections lined with acoustically absorbent materials and their experimental results demonstrate the effectiveness of a perforated steel plate, wire mesh and steel wool in attenuating C-J detonation [13].

The C-J deflagration phenomena were also observed in numerical simulations when detonation propagates in a tube with obstacles [17-24]. Heidari et al. developed a numerical approach to simulate flame acceleration and DDT in hydrogen-air mixture [17]. The numerical and theoretical study of Valiev et al. showed that the velocity of the accelerating flame saturates to a constant state, which can be correlated to the C-J deflagration [18]. Gamezo et al. conducted numerical simulations of DDT in tubes with obstacles by using the compressible reactive Navier–Stokes equations [22-24]. They found two mechanisms of DDT, one is DDT occurring in a gradient of reactivity and the other is from energy focusing as shock waves converge.

The experimental observations indicate that the structure and thermodynamics of C-J deflagration are totally different from that of C-J detonation. Lee thought that this combustion regime is a thermal choking regime [4-9]. Its front has a double-discontinuity structure, which is constituted by a leading shock wave and a chemical reaction zone. But the distance between the leading shock wave and the reaction zone is much wider than the induction zone of C-J detonation. Therefore, its thermodynamic properties cannot be interpreted simply by the classical C-J detonation theory.

Zhu et al. analyzed the thermodynamic properties of C-J deflagration experimentally and theoretically [15,16]. They found that it is not of the C-J type, being completely different from C-J detonation. Their experiments confirmed that the thermodynamic states of the reactant between the leading shock wave and the flame front are not the thermodynamic properties induced by the leading shock wave. The states in this region are influenced by both the original reactant and the combustion product.

Although C-J deflagration is usually obtained in three-dimensional obstacle-filled tubes, it has general one-dimensional properties. The mechanism of C-J deflagration is unknown to date; therefore, it becomes a research hotspot of fundamental detonation physics. In this paper, the mechanism of C-J deflagration is studied by using one-dimensional numerical simulation with overall one-step chemical reaction kinetics. The detonable mixture is the stochiometric $H_2$-air mixture at 1atm and 300K. The activation energy is increased in different cases to trigger the instability of detonation. New and original mechanisms about von Neumann spike, detonation instability, detonation quenching, and C-J deflagration propagation are revealed.

**2. Governing equations and physical model**

The governing equations for detonation simulation are one-dimensional Euler equations with the overall one-step chemical reaction kinetics. The viscosity, diffusion and heat conduction are neglected.

$$\frac{\partial U}{\partial t}+\frac{\partial F}{\partial x}=S \tag{1}$$

$$U = \begin{pmatrix} \rho \\ \rho u \\ \rho e \\ \rho Z \end{pmatrix} \quad F = \begin{pmatrix} \rho u \\ \rho u^2 + p \\ (\rho e + p)u \\ \rho u Z \end{pmatrix} \quad S = \begin{pmatrix} 0 \\ 0 \\ 0 \\ \dot{\omega} \end{pmatrix} \tag{2}$$

$$e = \frac{p}{(\gamma - 1)\rho} + \frac{1}{2}u^2 + Zq \tag{3}$$

$$\dot{\omega} = -K\rho Z \exp\left(-\frac{CE_a}{RT}\right) \tag{4}$$

where $\rho$, $p$, $u$, $e$, $q$ and $Z$ are density, pressure, velocity, specific energy, specific heat release, and chemical reaction progress parameter, respectively. $\dot{\omega}$ is the mass production rate of combustion product, $K$ is the pre-exponential factor, $T$ is the temperature in Kelvin $K$, and $E_a$ is the activation energy per unit mass of reactant, respectively. The parameter $C$ in Eq.(4) is a parameter used to change the activation energy.

The gas constant $R$ and the special heat ratio $\gamma$ are calculated as

$$R(Z) = R_1 Z + R_2(1 - Z) \tag{5}$$

$$\gamma(Z) = \frac{\gamma_1 R_1 Z / (\gamma_1 - 1) + \gamma_2 R_2 (1 - Z) / (\gamma_2 - 1)}{R_1 Z / (\gamma_1 - 1) + R_2 (1 - Z) / (\gamma_2 - 1)} \tag{6}$$

where the subscripts 1 and 2 represent reactant and product, respectively.

## 3. Numerical methods and conditions

One-dimensional numerical simulations were conducted to simulate the C-J deflagration. The computational domain is a straight detonation tube with the left end closed and the right end open. The detonation tube is fully filled with the premixed stoichiometric $H_2$-Air mixture at 1 atm and 300 K. Detonation is initiated by a smaller region of reactant with high pressure and high temperature specified near the closed end, and then it propagates from left to right.

Two-order ENO scheme and three-order TVD Runge-Kutta method are applied [25]. The convective flux is split by Steger-Warming flux splitting method [26]. The mirror reflection boundary conditions are applied on the closed end. Grid resolution study is conducted to eliminate the influence of grid sizes. The uniform grid spacing are *dx*= 0.1mm and 0.05 mm, respectively. The parameters of this overall one-step detonation model for stoichiometric $H_2$-Air mixture at initial conditions of 1atm and 300K are $Z_1 = 1.0$ , $Z_2 = 0$ , $\gamma_1 = 1.40$ , $\gamma_2 = 1.29$ , $R = 398.5\text{J/(kg}\cdot\text{K)}$ , $R = 368.9\text{J/(kg}\cdot\text{K)}$ , $q = 3.5\times10^6\,\text{J/kg}$ , $E_a = 4.794\times10^6\,\text{J/kg}$ , $K = 7.5\times10^9\,\text{s}^{-1}$ . It was demonstrated that this model can simulate the detonation cell size quantitatively [27, 28].

## 4. Results and discussion

### 4.1 Detonation to deflagration transition

The parameter C is assumed to be a constant in each case and its value is increased to increase the activation energy in different cases. Figure 1 shows the detonation propagation speed with different values of parameter C from 1.0 to 1.567. The x-coordinate is time from the start of calculation and the y-coordinate is detonation wave speed. The uniform grid space is *dx*=0.1mm. For C=1.0, the result is C-J detonation with a theoretical propagation speed of 1950 m/s. As the value of C increases from 1.0 to 1.566, the detonation speed keeps the theoretical value of C-J detonation speed.

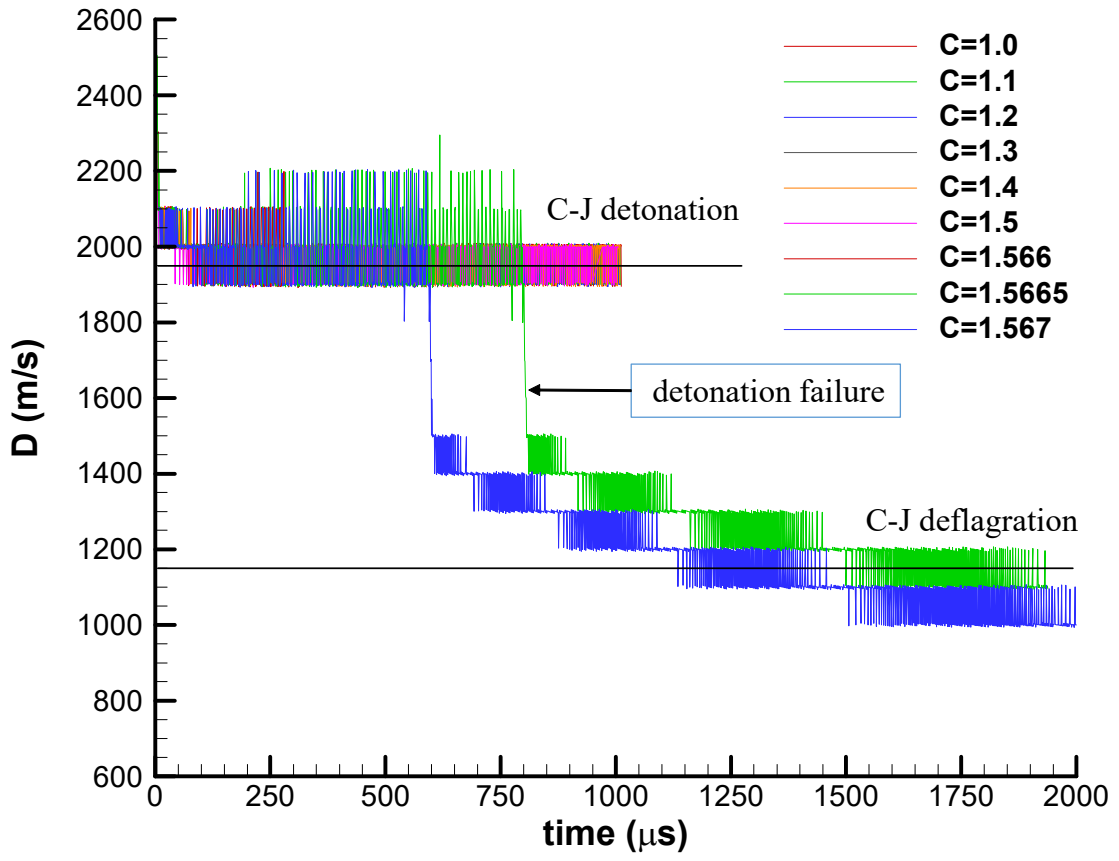

Fig.1 Detonation and deflagration speed with different activation energy

When the parameter C is increased to 1.5665, the detonation speed decreases to about 1150 m/s abruptly and then keeps this speed almost constantly for a long time, which means that an abrupt change of the flow field occurs. The Mach number of the C-J deflagration is Ma=2.8 and its propagation speed is about 58% $D_{C\text{-}J}$. This is a critical condition for DDT process. The C-J deflagration cannot maintain the initial propagation speed of about half $D_{C\text{-}J}$ and is observed to decay rapidly. The onset of detonation is never observed for the remainder of the tube. The grid resolution study was conducted and the uniform grid spacing was reduced to be dx=0.05mm. The detonation speeds at different parameter C are the same, demonstrating that the results are independent of grid size.

The pressure profiles of C-J detonation with different activation energy with the time interval of 20 μs are plotted in Fig.2 for comparison. The x-coordinate is the length of detonation tube and the y-coordinate is the pressure ratio. The detonation propagates from left to right and x=0 is the end wall. Firstly, we can find that the profiles of C-J detonation with different activation energy are similar, which means that their flow fields are the same. Secondly, their C-J pressure ratios are 15, being equal to the theoretical value of C-J detonation. Thirdly, the positions of zero-velocity point are the same and the pressure ratio at wall is 5.8. Finally, the key difference is that the C-J detonation with C=1.0 does not have the von Neumann spike. The others with C>1.0 have von Neumann spike and the value of von Neumann spike increases with the increase of activation energy. The key mechanism of von Neumann spike on the instability of detonation will be discussed in the following subsection.

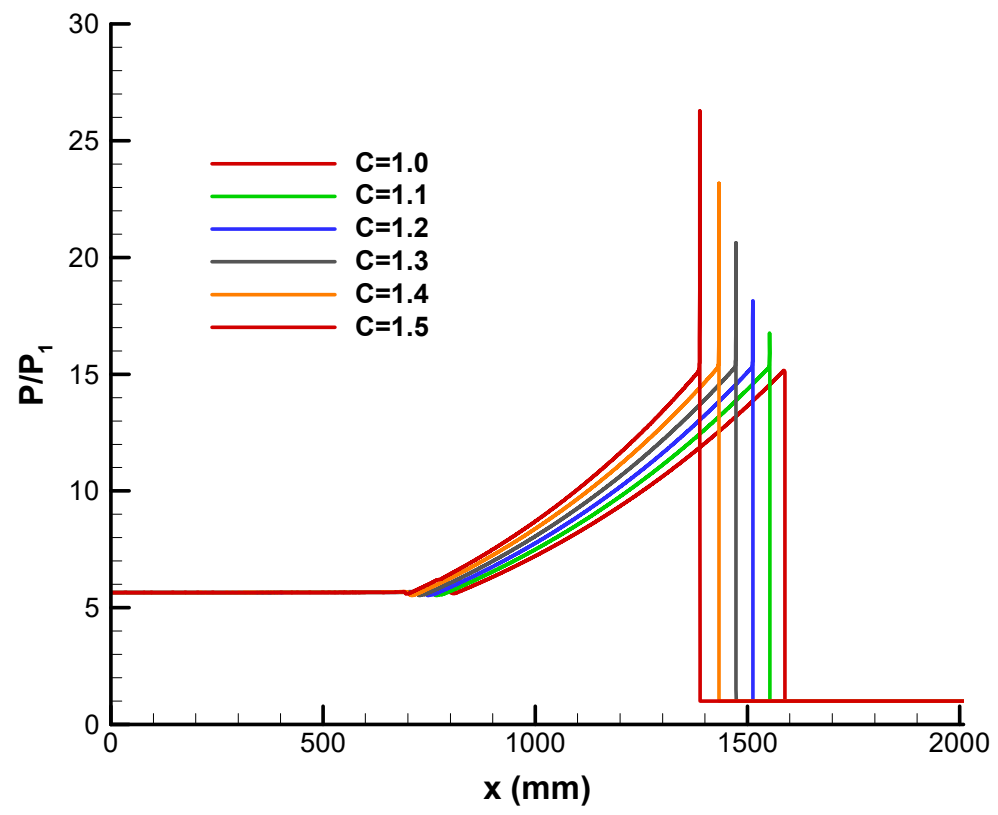

Fig.2 Pressure profiles of C-J detonation with different activation energy

The pressure profile of C-J deflagration at time 1000 μs is shown in Fig.3. The mass fraction of reactant Z is plotted to show the position of flame front (Fig.3a). It is also compared with experimental results of reference [15,16] qualitatively. The x-coordinate is turned from right to left in order to facilitate the comparison (Fig.3b). It is seen that the pressure profile of C-J deflagration is completely different from that of C-J detonation. The structure of the deflagration front consists of a leading shock wave followed by a flame and the thickness between them is lengthening with time. The pressure ratio across the precursor shock is measured to be about 12 at this instant, which is about the pressure ratio behind a normal shock wave propagating at Mach 2.8. The flame front is completely separated from the precursor and the pressure drops rapidly behind the reaction front.

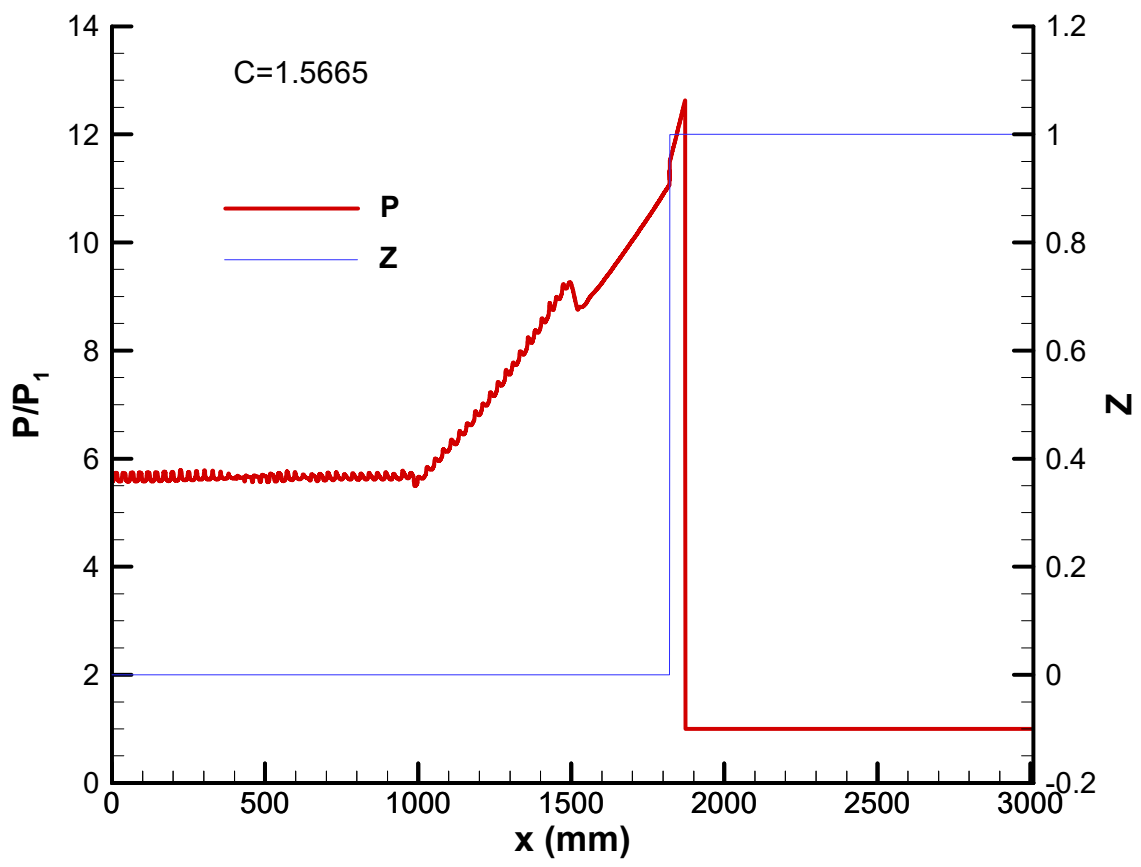

(a) Pressure profile of C-J deflagration

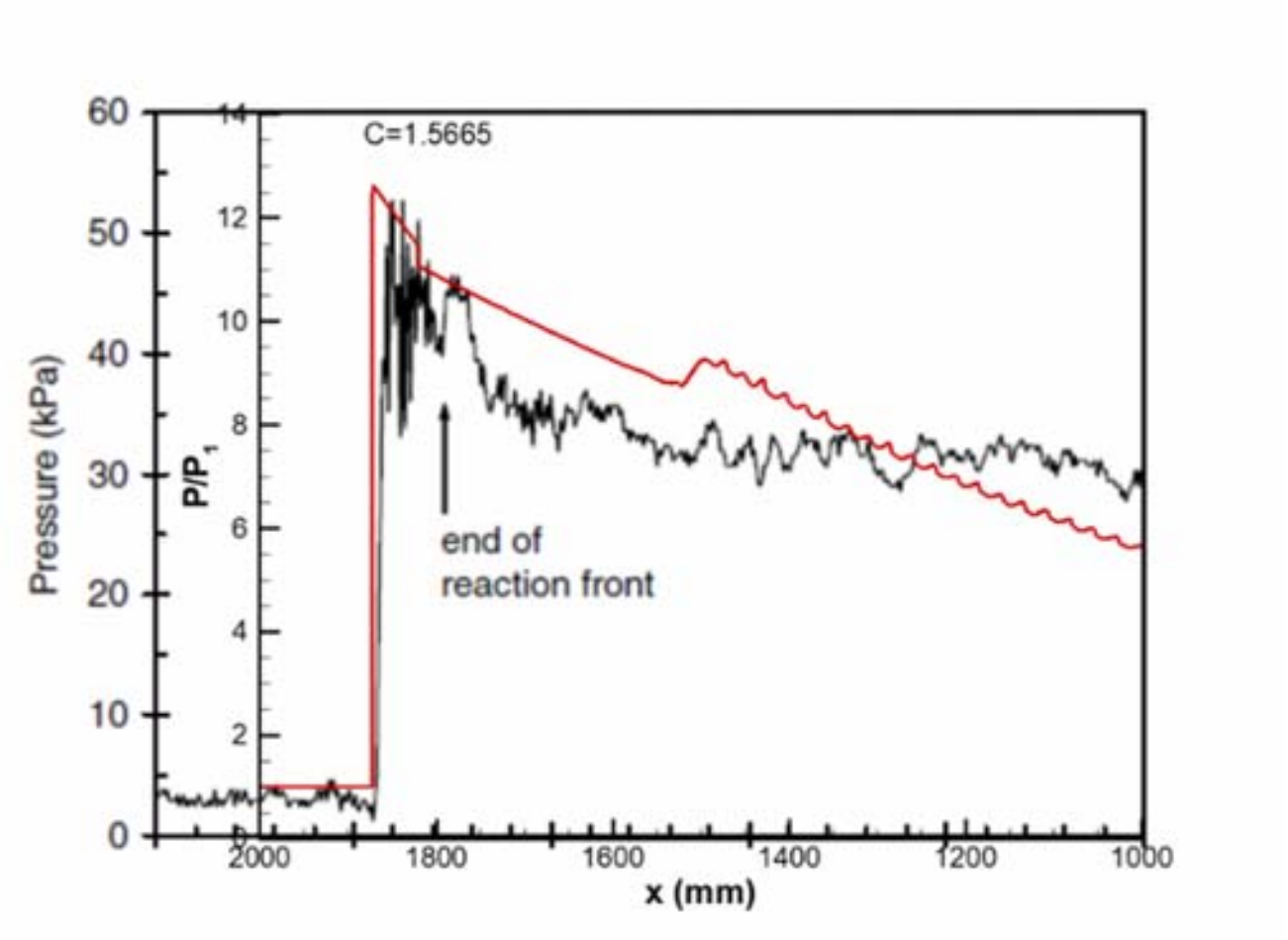

(b) Comparison with experimental results [15,16]

Fig.3 Pressure profile of C-J deflagration and its comparison with experiments

### 4.2 The Instability of Detonation

The C-J detonation goes the process of stable propagation, unstable propagation and quenching with the increase of activation energy. Then, it degenerates into a C-J deflagration. The profiles of gas

Mach number in the laboratory coordinate at the instants corresponding to Fig.2 are plotted in Fig.4. We can find that the gas Mach number at the detonation front rises from subsonic to supersonic with the increase of activation energy. Theoretically, for a normal shock wave with Mach number 4.8, it induces supersonic flow with Mach number 1.78 behind it, and the gas temperature and velocity are 1523K and 1596m/s, respectively. If the heat release process is very fast, the product velocity and Mach number behind a C-J detonation decreases to 849 m/s and Mach number 0.75, respectively. However, if the heat release process is slower, the detonation front maintains supersonic according to the ZND model.

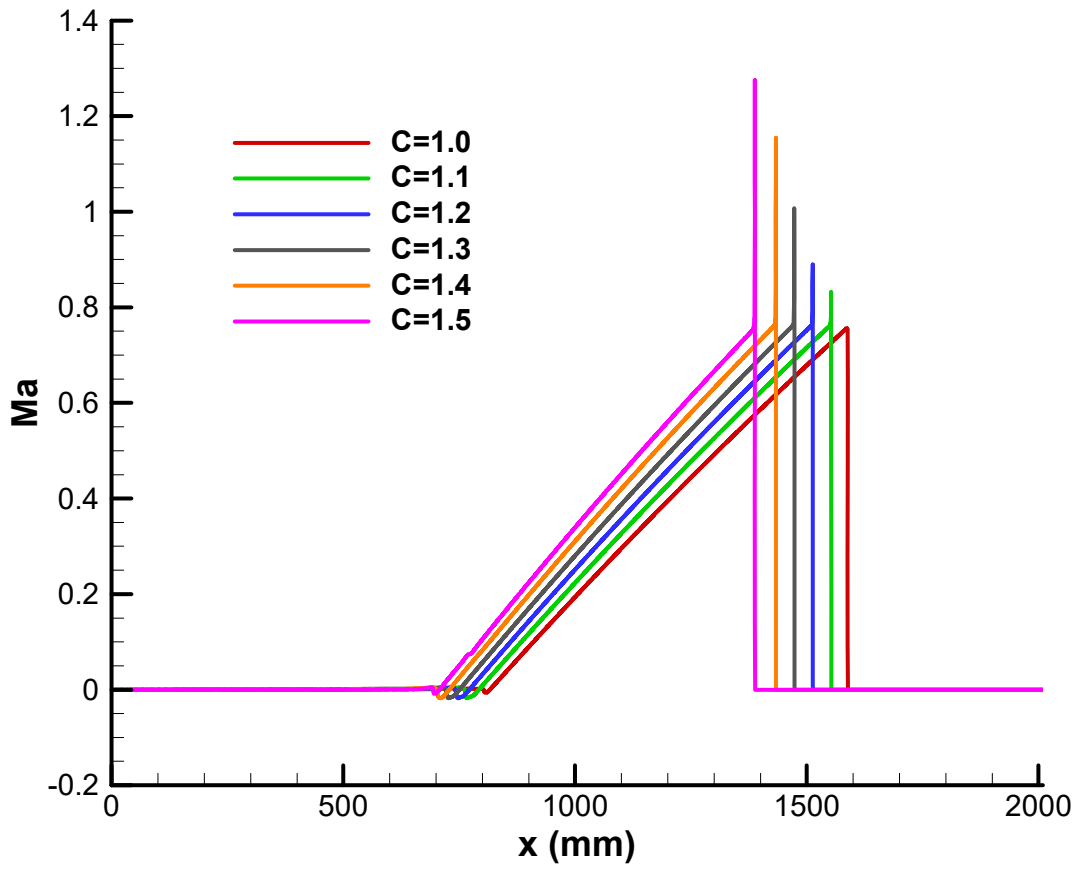

Fig.4 Mach number profiles of C-J detonation with different activation energy

In this numerical simulation, for the C-J detonation with C=1.0, the maximum Mach number at detonation front is 0.75, being equal to the theoretical value. The maximum Mach number at von Neumann spike becomes 1.0 for C=1.3 and 1.26 for C=1.5, respectively. The maximum pressure spike and maximum Mach number spike at detonation front with different activation energy are recorded in Fig.5 and Fig.6, respectively. We find that the detonation instability appears with the von Neumann spike synchronously. For a C-J detonation without von Neumann spike, it is stable. The higher the von Neumann spike is, the more unstable the detonation will be.

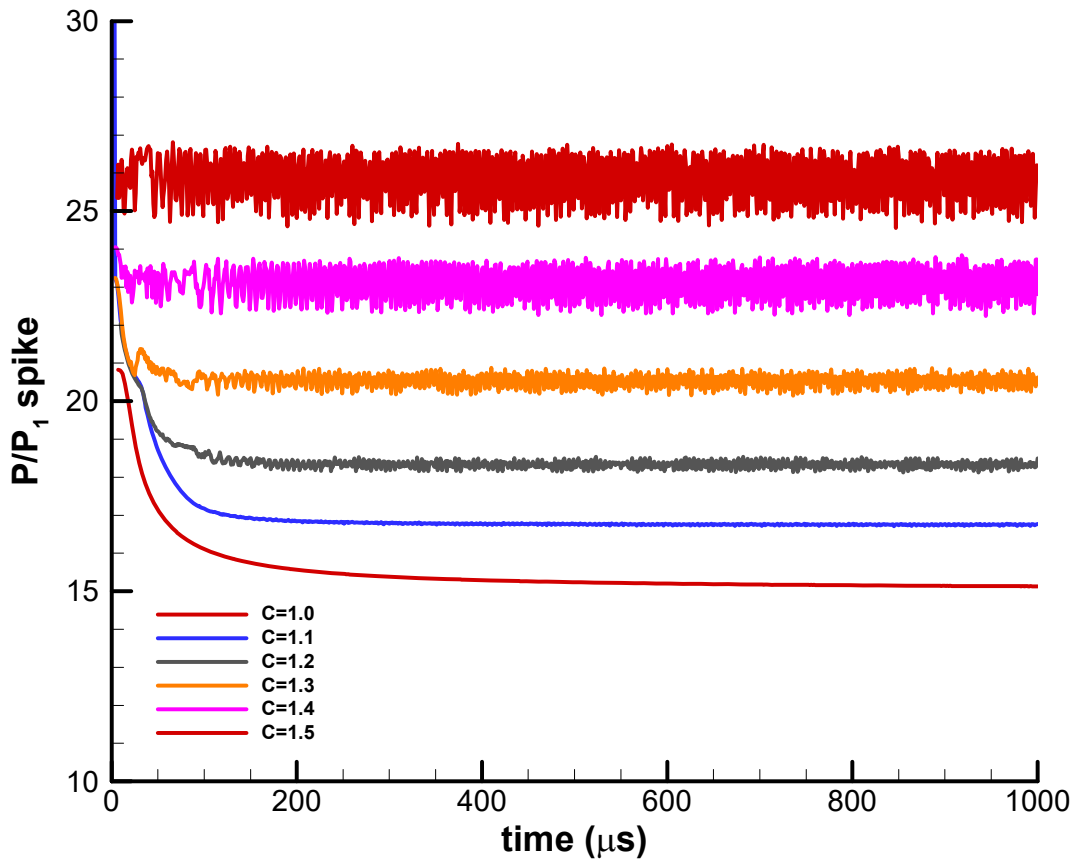

Fig.5 The maximum pressure at detonation front with different activation energy

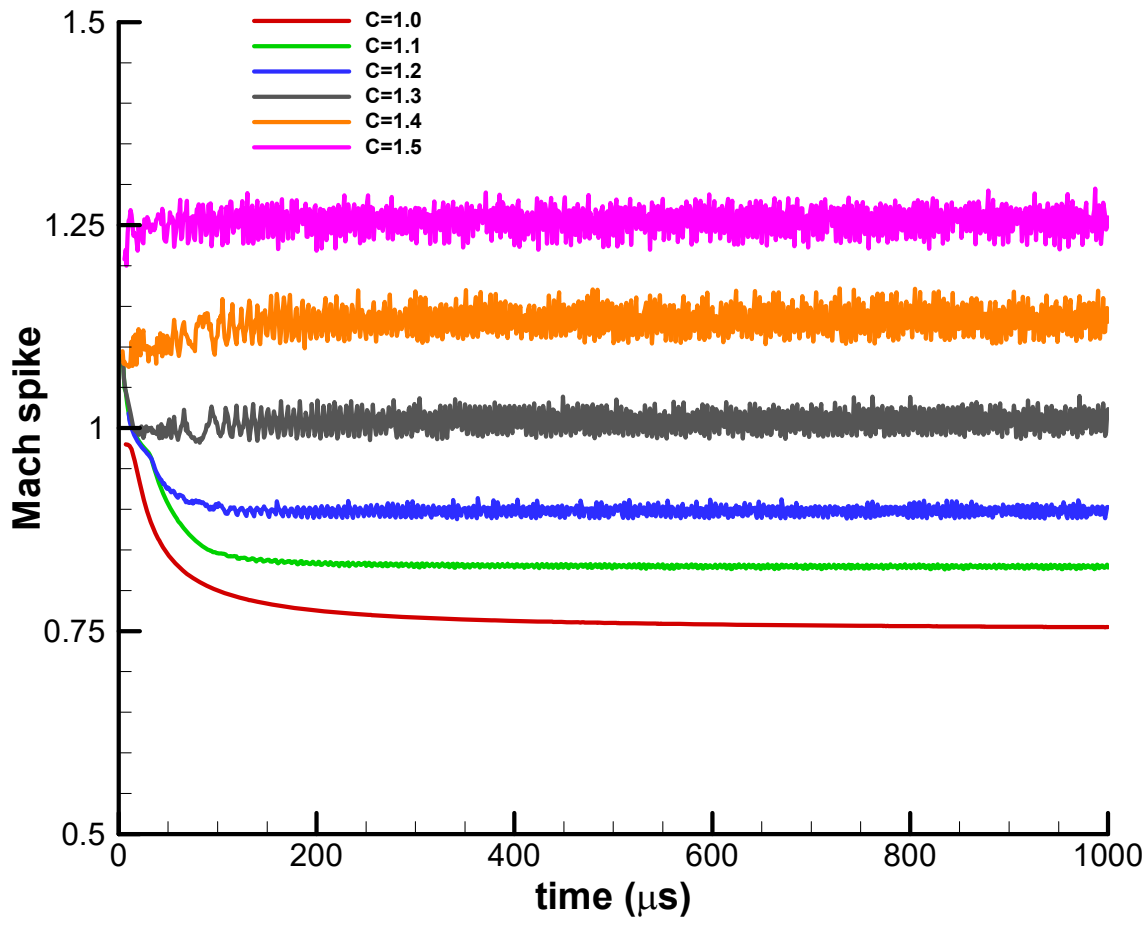

Fig.6 The maximum gas Mach number at detonation front with different activation energy

The mechanism of detonation instability can be explained by the classical linear stability theory, which is back to Rayleigh [29]. According to Rayleigh's theory, the velocity profile with an inflectional point is unstable for inviscid flow. For a detonation with a von Neumann spike, there indeed an inflectional point behind the detonation front. In addition, according to Eq. (1), $U^{n+1}=U^{n}-\frac{\partial F}{\partial x}\Delta t$. If there is an inflectional point at the detonation front, the derivative of convective flux is not a constant, which will induce disturbance to the flow field. The higher von Neumann spike means stronger rarefaction waves. For the TVD-type upwind numerical scheme, the convective flux is split into three parts, $u$, $u-a$, $u+a$. If the flow in the detonation front is supersonic, all the fluxes are constructed upwind, that is to the left wall direction or the rarefaction wave direction. If the detonation front is subsonic, $u-a<0$, a part of the flux is constructed downstream or to the shock wave direction.

$$F = F_1 + F_2 + F_3 \tag{7}$$

$$F_1 = \frac{\rho(u-a)}{2\gamma}\begin{pmatrix}1\\ u-a\\ h-ua\\ Z\end{pmatrix}, F_2 = \frac{(\gamma-1)\rho u}{\gamma}\begin{pmatrix}1\\ u\\ \frac{1}{2}u^2+Zq\\ Z\end{pmatrix}, F_3 = \frac{\rho(u+a)}{2\gamma}\begin{pmatrix}1\\ u+a\\ h+ua\\ Z\end{pmatrix} \tag{8}$$

As the activation energy is increased further, the whole pressure profiles of detonation become unstable as shown in Fig.7, which is also the omen of detonation quenching. Figure 8 shows the quenching process of unstable detonation. The time interval between two successive profiles is 10μs. We can find that the unstable detonation propagates downstream at first and then dies out abruptly after the last oscillation. It is extinguished only once time by the very stronger rarefaction wave generated by the higher von Neumann spike.

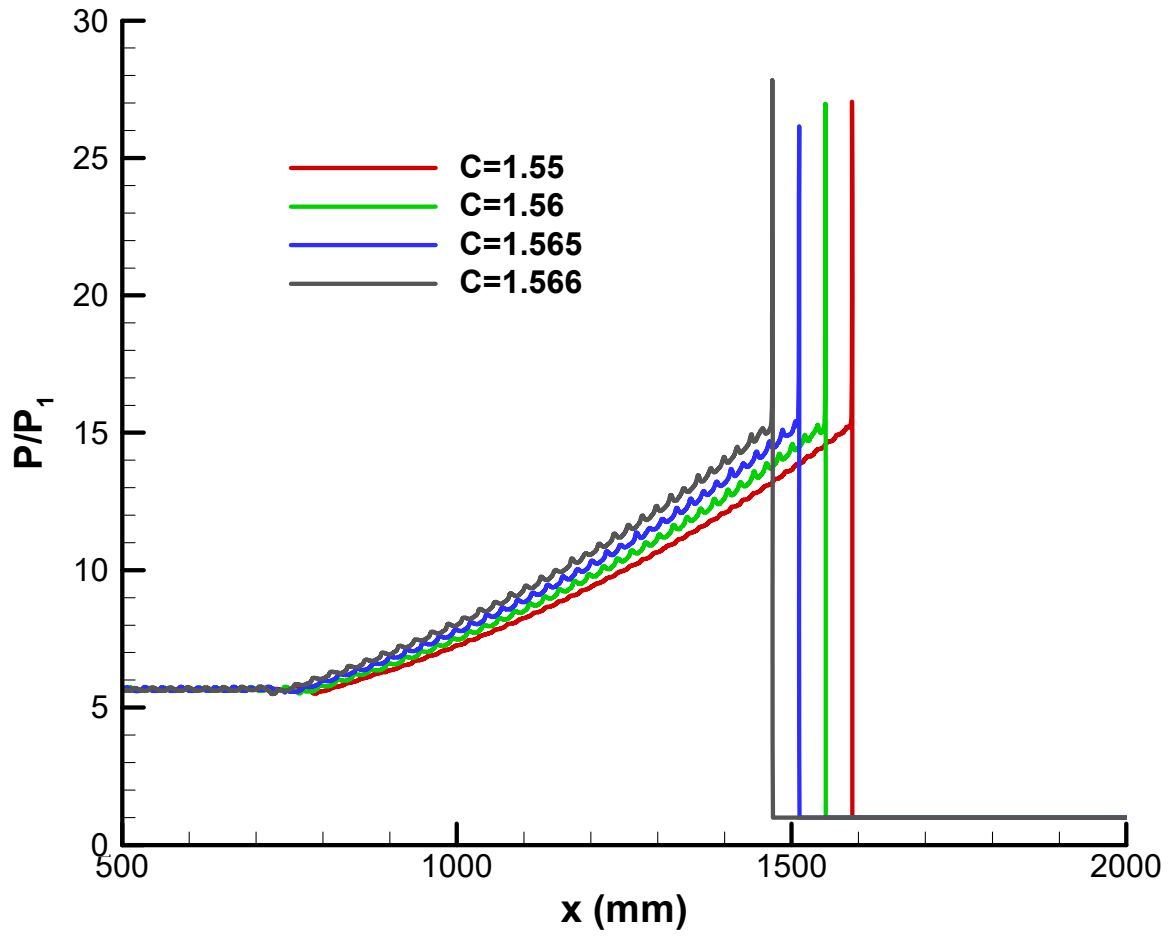

Fig.7 The pressure profiles of unstable detonation at higher activation energy

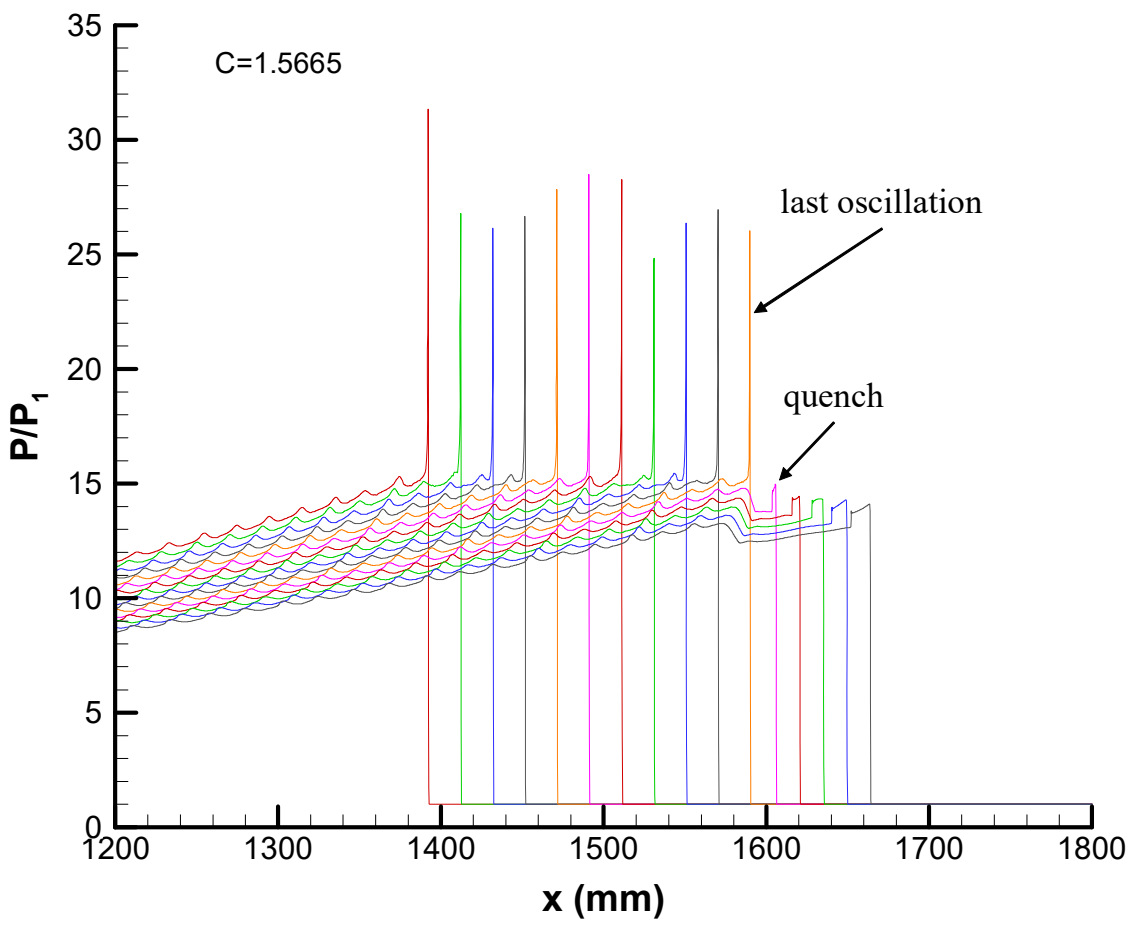

Fig.8 The quenching process of unstable detonation at higher activation energy

In order to discuss the mechanism of detonation quenching, the convective flux of energy is discussed. Define the convective flux of energy equation as $\Gamma = -\frac{\partial(\rho e + p)u}{\partial x}dt$ . In numerical simulations, at each time step, $\Gamma > 0$ means compression wave and $\Gamma < 0$ means rarefaction wave. The Energy convective flux and heat release $\Delta Q$ of C-J detonation with C=1.0 are plotted in Fig.9. We can find that the convective flux and the heat release are coupled with each other tightly. The rarefaction wave is too weak to be seen.

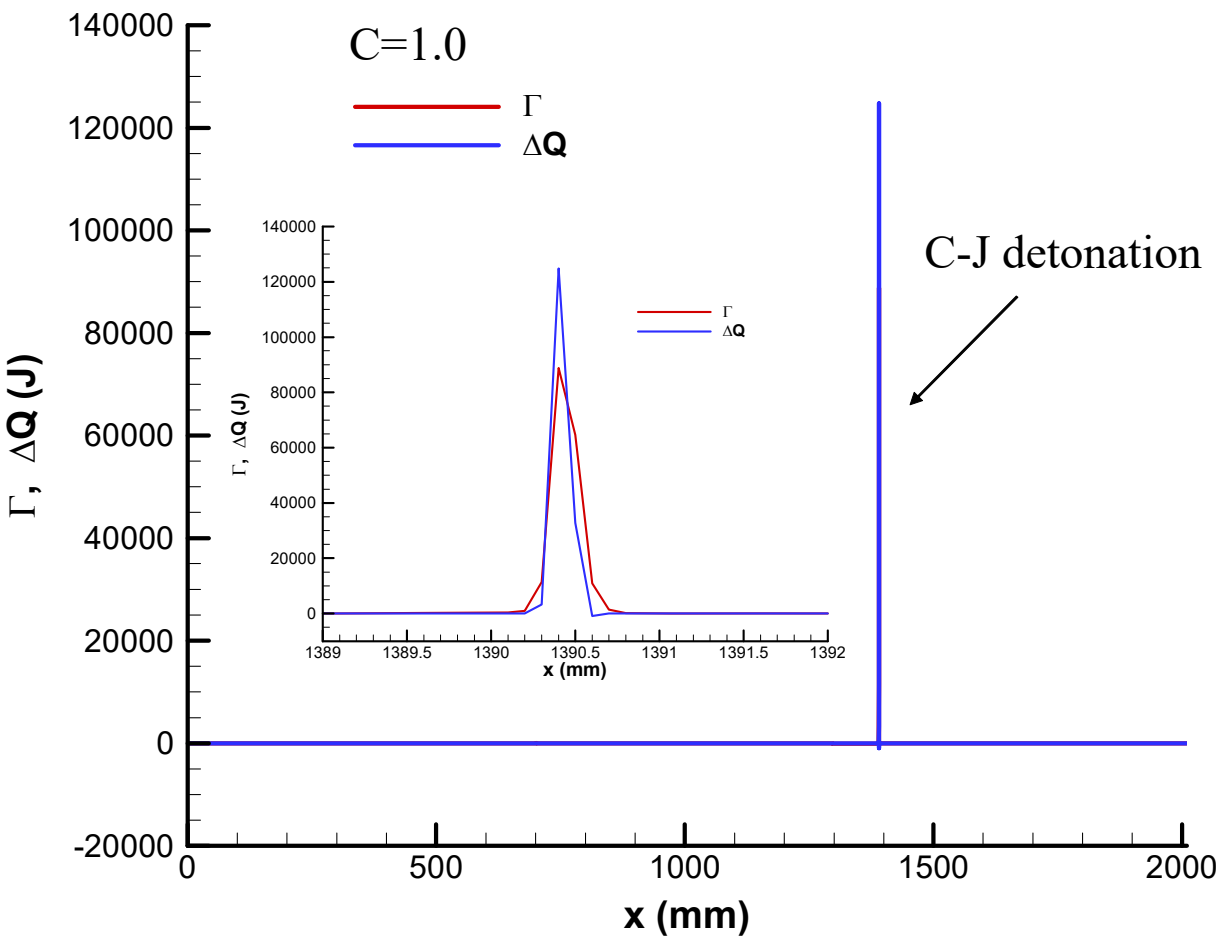

Fig.9 The Energy flux and heat release of C-J detonation with C=1.0

The convective flux and heat release during the unstable detonation quenching process are plotted in Fig.10. We find that at the critical activation energy case, the strength of rarefaction wave is stronger than the heat release. For the unstable detonation just before quenching, the rarefaction wave is behind the flame front (Fig.10a). For the detonation at the critical instant of quenching, the rarefaction wave and the flame front are at the same position (Fig.10b) and extinguishes the detonation. The pressure profile just at the instant of quenching corresponding to Fig.10b at t=801μs is plotted in Fig.11. We can see that there is pressure concavity at the flame front. Just after the quenching, the rarefaction wave is in front of the flame front and weakens the precursor shock wave gradually to a quasi-state of C-J deflagration.

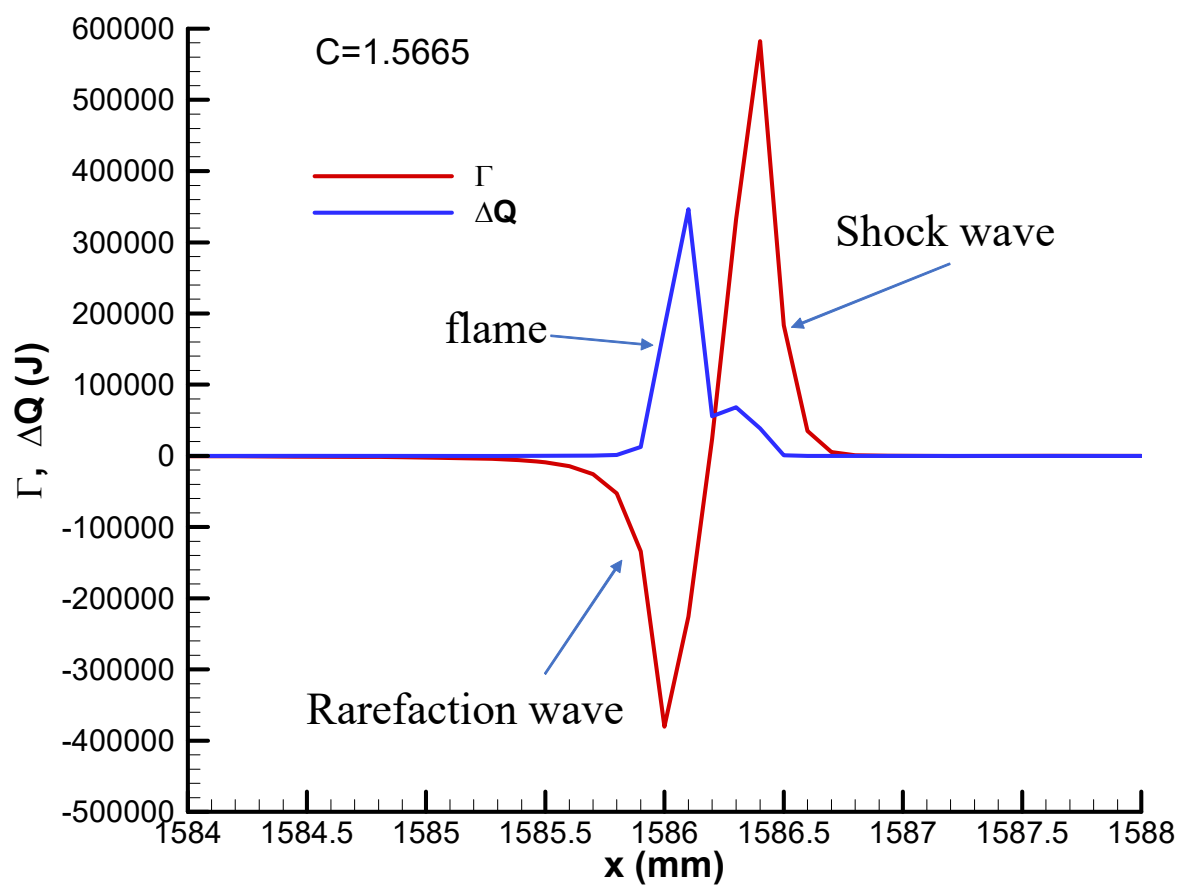

(a) the convective flux just before quenching (t=798μs)

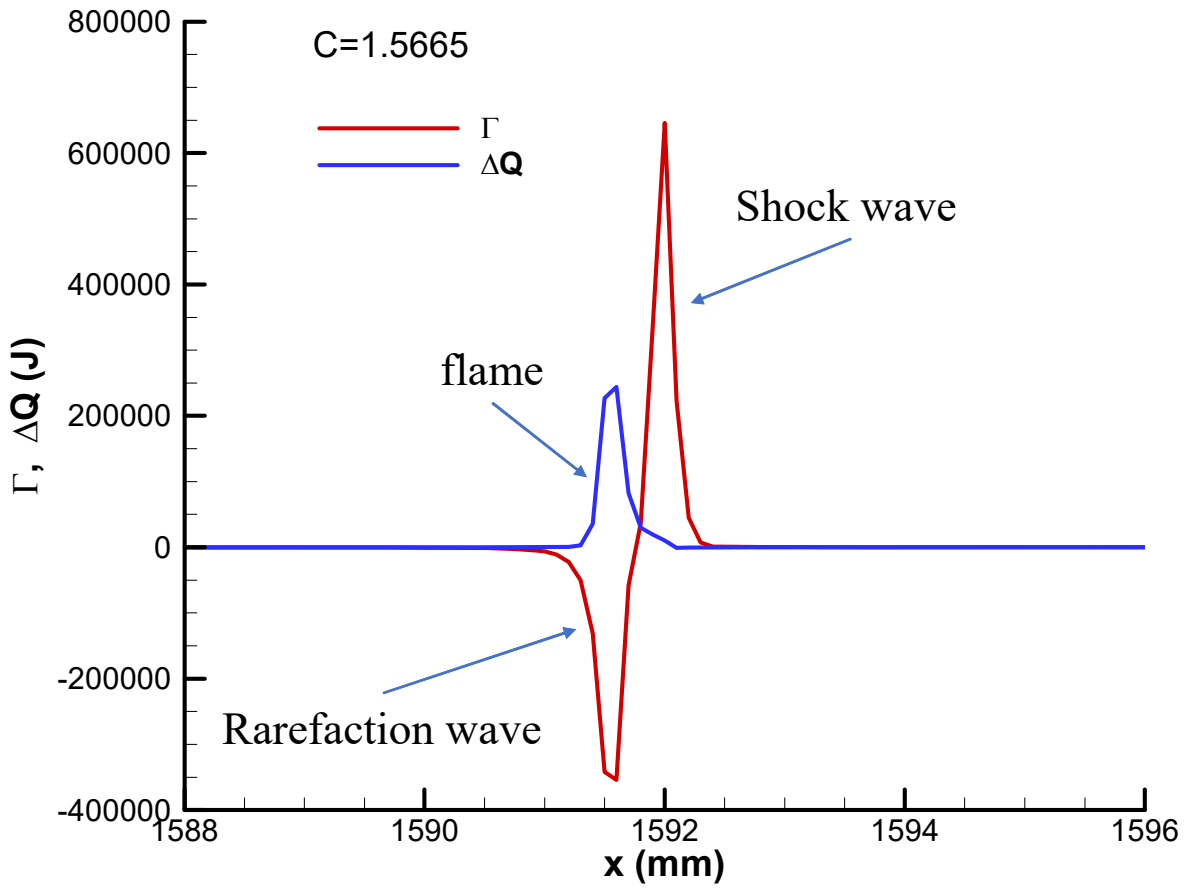

(b) the convective flux at the critical instant of quenching (t=801μs)

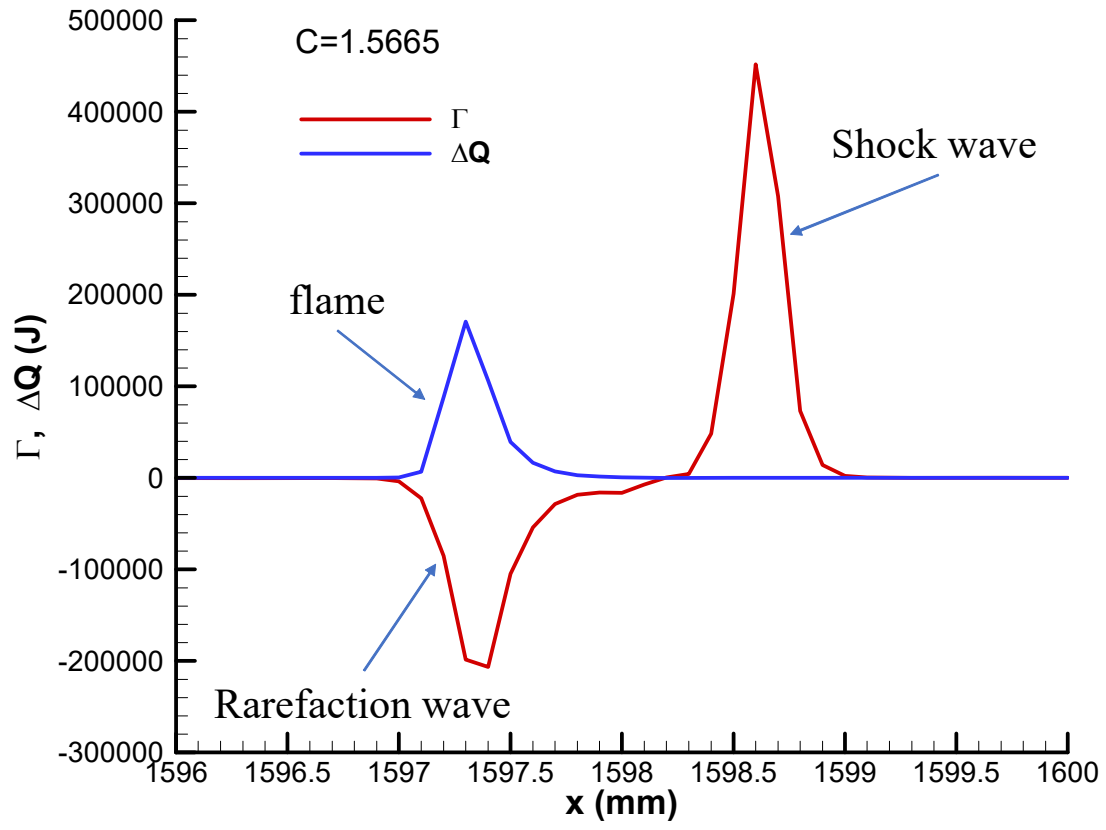

(c) the convective flux just after quenching (t=805μs)

Fig.10 The convective flux and heat release during the detonation quenching process

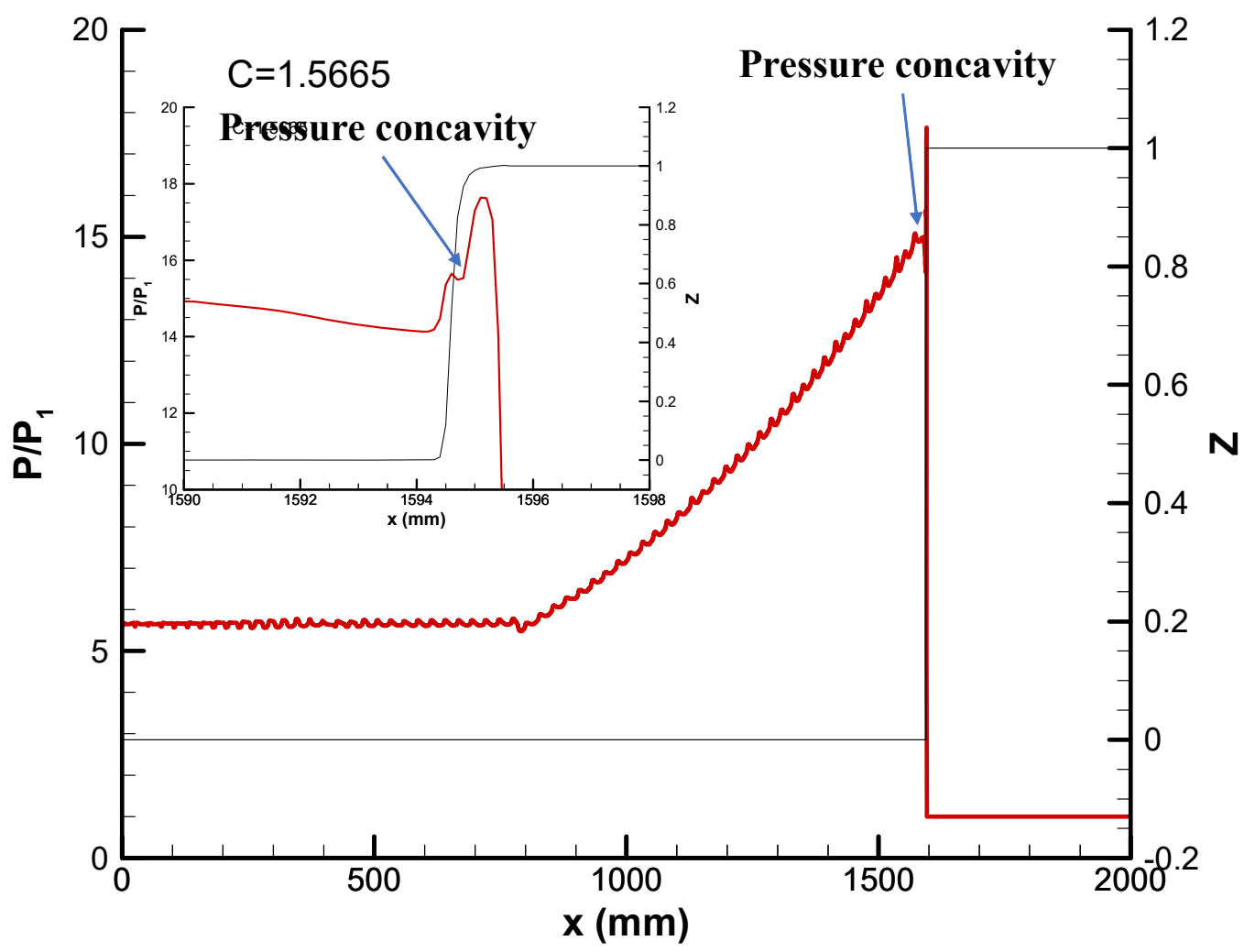

Fig.11 The pressure profile just at the instant of quenching

### 4.3 The Mechanism of C-J deflagration

The C-J deflagration does not propagate downstream with a constant velocity. Figure 12 shows a sequence of deflagration pressure profiles. The x-t diagram of C-J deflagration and its comparison with experimental results [15,16] are plotted in Fig.13. The numerical results agree with experimental results very well. The unstable detonation propagates at first and then quenches suddenly. The C-J deflagration is observed to decay rapidly and the thickness between the precursor shock wave and the flame front increases as it propagates downstream. The flame is found to propagate at a relatively constant velocity that is only about 10% lower than that of the precursor shock wave and the slight lengthening of the reaction zone can be seen in the figure.

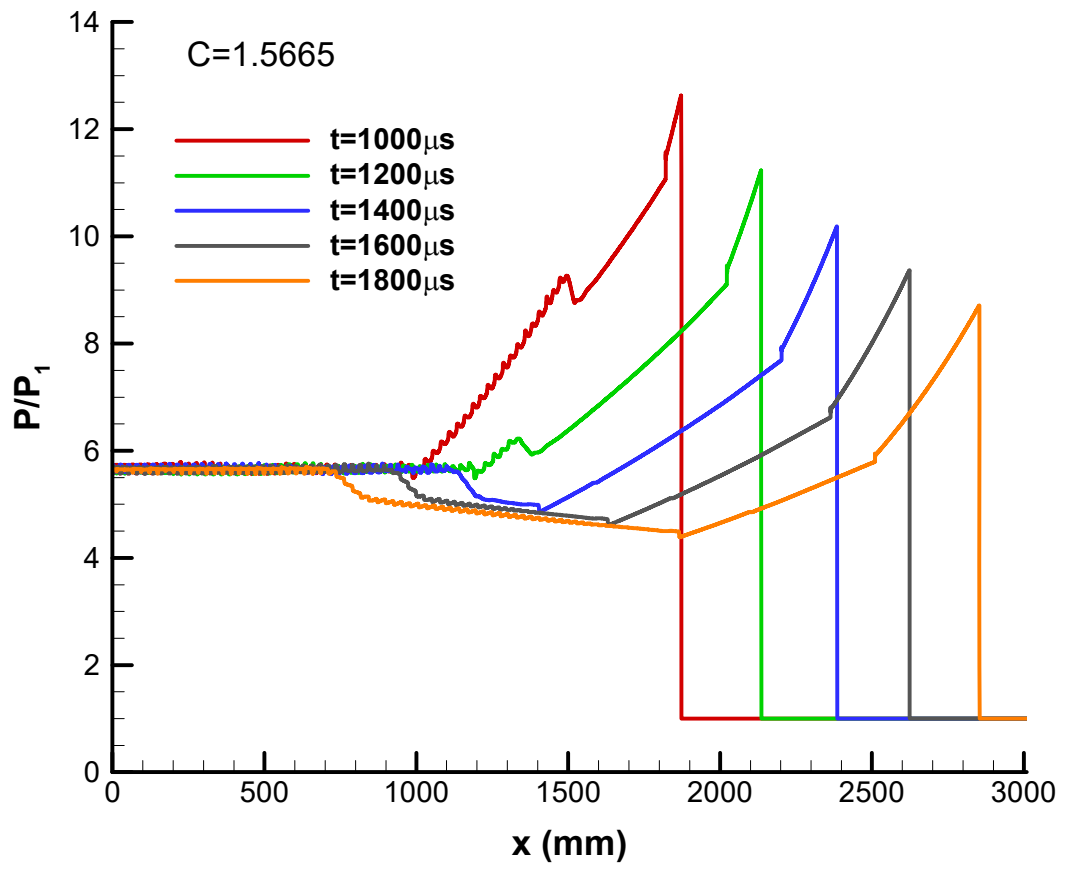

Fig.12 The propagation process of C-J deflagration at difference time

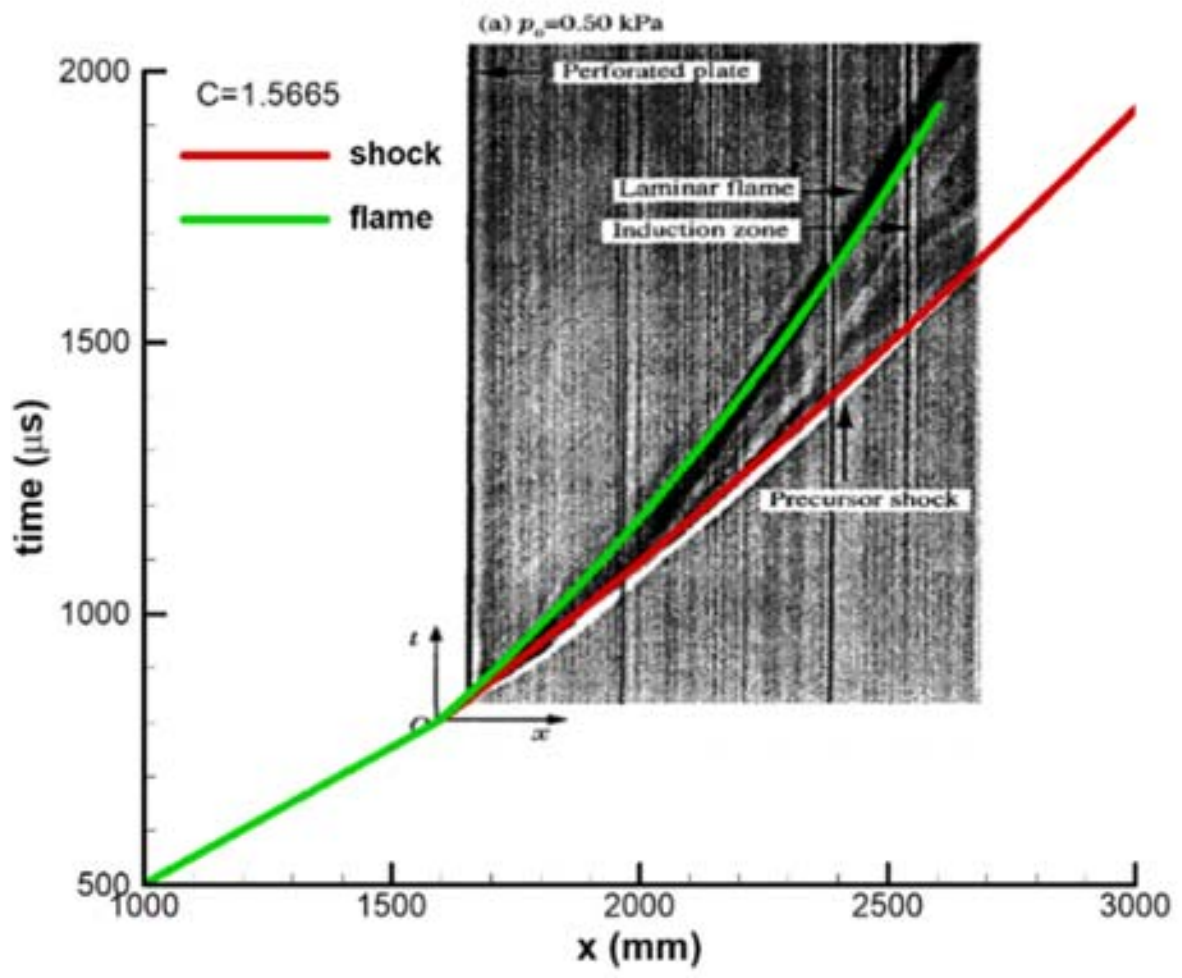

Fig.13 The x-t diagram of C-J deflagration and comparison with experiments [15,16]

Theoretically, for a normal shock wave propagating with a Mach number of 2.8, the static temperature, velocity, and gas Mach number behind the shock wave is 726K, 838m/s and Ma1.33, respectively. The static temperature behind the leading shock wave is too low to ignite the detonable mixture. The structure of the deflagration front is shown in Fig.14. The frozen pressure profile without heat release and pressure profile after heat release at one time step in numerical simulation are compared. We can find that the deflagration front consists of a precursor shock wave and a flame front. The rarefaction waves cease the velocity of gas behind the precursor shock.

Figure 15 shows the velocity and Mach number profiles of C-J deflagration in the laboratory coordinate. The velocity is decreased from 1000m/s to 800m/s and the Mach number is decreased from 1.3 to 0.8 at the flame front by rarefaction waves. As a result, its static temperature is increased and its static pressure is reduced. The constant-volume combustion occurs at the interface synchronously and increase the pressure to its original value, offsetting the influence of rarefaction waves. The propagation velocity of flame is the sound velocity of detonation product; therefore, if the propagation velocity of the precursor shock is equal to the sound velocity of detonation product, the C-J deflagration will keep a quasi-steady state. The sound velocity of the detonation product of stoichiometric $H_2$-air mixture is 1130 m/s, being about 57% of detonation velocity. This is the key mechanism which drives the C-J deflagration propagating with a relatively constant velocity for some distance.

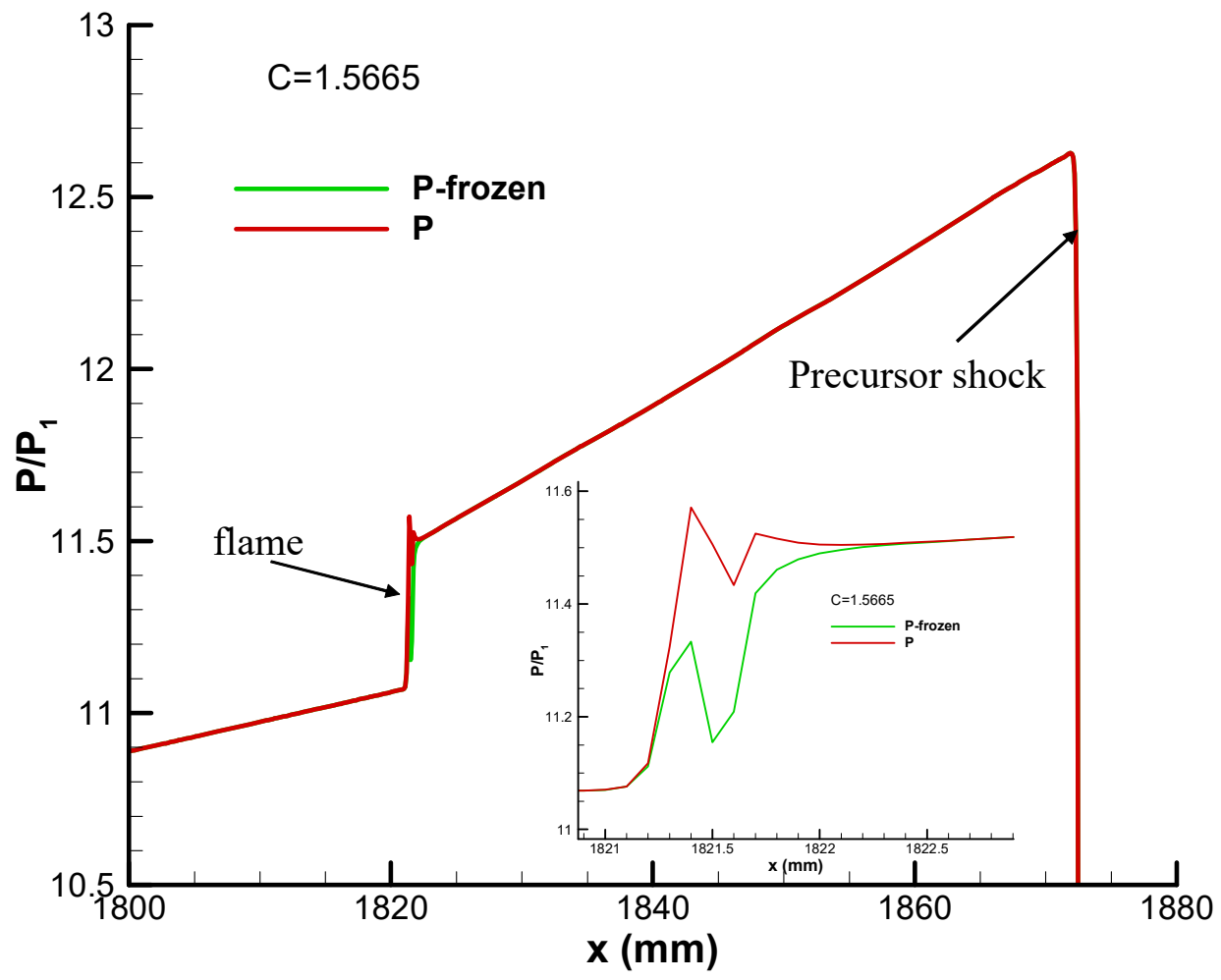

Fig.14 The structure of C-J deflagration front

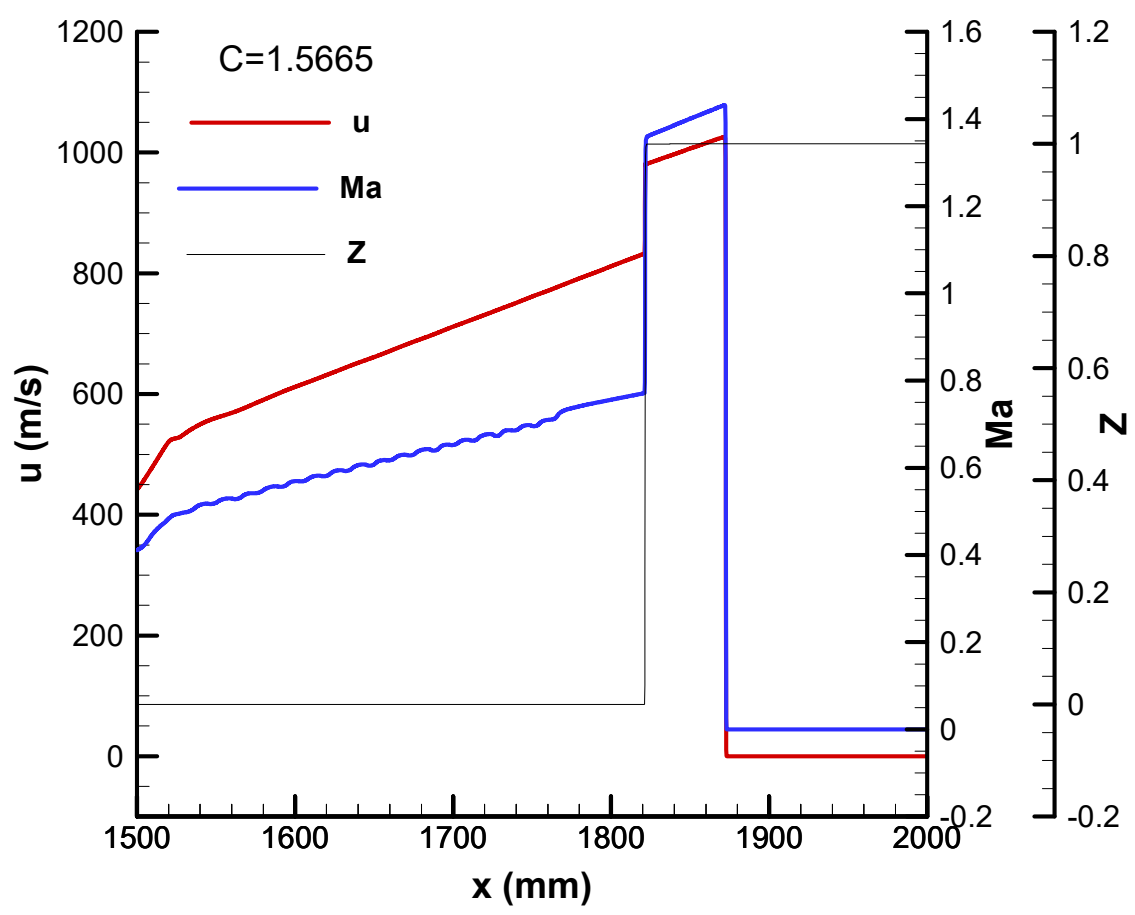

Fig.15 The velocity and Mach number profiles of C-J deflagration

## 5. Conclusions

In this paper, the mechanism of detonation instability and C-J deflagration propagation are studied by using one-dimensional numerical simulation and overall one-step chemical reaction kinetics. The activation energy is increased to induce the instability of detonation. This numerical study comes to the following conclusions.

(1) The C-J detonation does not have von Neumann spike at lower activation energy, but has von Neumann spike at higher activation energy. The peak value of von Neumann spike increases as the activation energy increases. The gas Mach number at detonation front is subsonic at lower activation energy and becomes supersonic at higher activation energy. The theoretical value of gas Mach number for a C-J detonation is 0.75. The higher activation energy leads to higher gas Mach number at detonation front.

(2) The C-J detonation instability appears with the von Neumann spike synchronously. The higher the von Neumann spike is, the stronger the instability will be. The mechanism of detonation instability can be explained by the linear stability theory of Rayleigh, which is that the velocity profile with an inflectional point is unstable for inviscid flow. The quenching of detonation will occur when it becomes more unstable. The higher von Neumann spike means stronger rarefaction waves. The detonation is extinguished by the rarefaction wave abruptly, which is stronger than the heat release. After that, the rarefaction wave moves in front of the heat release region and the shock wave decays gradually.

(3) The C-J deflagration is composed of a precursor shock wave and a flame front. The flame front is completely decoupled from the leading shock wave. The static temperature of gas induced by the leading shock wave is too low to auto-ignite. The rarefaction wave from the left wall ceases the mixture behind the leading shock and increases its static temperature. As a result, the constant-volume combustion takes place at the interface automatically. The pressure gain generated by the combustion at the interface offsets the pressure decrease caused by the rarefaction wave. It is this key mechanism that makes the precursor shock wave propagate downstream with a relatively constant velocity for a certain distance.